\documentclass[aps,float,twocolumn,epsfig]{revtex4}
\usepackage{amsmath}
\usepackage{epsfig}

\begin{document}

\title{Analytical study of four-wave mixing with large atomic coherence}
\author{E.A. Korsunsky$^{(1)}$, T. Halfmann$^{(1)}$, J.P. Marangos$^{(2)}$, and K. Bergmann$^{(1)}$}

\affiliation{$^{(1)}$ Fachbereich Physik, Universit\"{a}t
Kaiserslautern, D-67663 Kaiserslautern, Germany\\
$^{(2)}$ Physics Department, Blackett Laboratory, Imperial
College, London SW7 2BZ, United Kingdom}

\date{\today{}}

\begin{abstract}
Four-wave mixing in resonant atomic vapors based on maximum coherence
induced by Stark-chirped rapid adiabatic passage (SCRAP) is investigated
theoretically. We show the advantages of a coupling scheme involving maximum
coherence and demonstrate how a large atomic coherence between a ground and
an highly excited state can be prepared by SCRAP. Full analytic solutions of
the field propagation problem taking into account pump field depletion are
derived. The solutions are obtained with the help of an Hamiltonian approach
which in the adiabatic limit permits to reduce the full set of Maxwell-Bloch
equations to simple canonical equations of Hamiltonian mechanics for the
field variables. It is found that the conversion efficiency reached is
largely enhanced if the phase mismatch induced by linear refraction is
compensated. A detailed analysis of the phase matching conditions shows,
however, that the phase mismatch contribution from the Kerr effect cannot be
compensated simultaneously with linear refraction contribution. Therefore,
the conversion efficiency in a coupling scheme involving maximum coherence
prepared by SCRAP is high, but not equal to unity.
\end{abstract}

\maketitle

\section{Introduction}

Nonlinear frequency conversion processes in atomic or molecular gases have
attracted much attention since the early days of nonlinear optics. The
interest is mainly motivated by the possibility to generate coherent
radiation in the XUV and VUV frequency range, where there are no transparent
nonlinear crystals. However, the conversion efficiencies are usually
relatively poor due to small nonlinear susceptibility for the generation and
difficulties to prepare proper phase matching conditions. Approaching atomic
resonances enhances the nonlinearity, but at the same time absorption,
linear refraction and unwanted nonlinear phase shifts increase rapidly.

Recently, a new technique has been put forward which substantially improves
the nonlinear-optical properties of a medium. The technique, usually
referred to as ''nonlinear optics with maximum coherence'', is based on the
preparation of all atoms in the medium in the same coherent superposition of
two states $\left| 1\right\rangle $ and $\left| 2\right\rangle $ with equal
probability amplitudes \cite{jain96}. From the classical point of view,
coherently prepared atoms represent an ensemble of dipoles all oscillating
at the same frequency $\omega _{21}$, with the same phase and with maximum
amplitude. If a radiation field of frequency $\omega _{3}$ is applied to the
medium, it will beat against this strong local oscillator to produce the
sum- or difference frequency $\omega _{21}\pm \omega _{3}$. In this case,
the nonlinear susceptibility of the generation process is large (in fact, it
is resonantly enhanced) and is of the same order as the linear
susceptibility. Therefore, complete conversion occurs within an optical
length smaller than the coherence length. Consequently, requirements for the
phase matching are substantially alleviated and the influence of
density-dependent detrimental effects is minimized.

In first proposals and experimental implementations \cite{jain96,mer99},
maximum coherence was established in a lambda-type coupling scheme with
ground $\left| 1\right\rangle $ and lower excited $\left| 2\right\rangle $
states using stimulated Raman adiabatic passage (STIRAP) \cite{stirap}.
However, in a lambda-type coupling scheme involving one-photon transitions
the generated radiation $\omega _{21}\pm \omega _{3}$ cannot reach far into
the vacuum-ultraviolet spectral region \cite{mer99}. Multi-photon
excitations may not be used in order to reach higher lying states, because
laser-induced Stark shifts, which are intrinsic to multi-photon transitions,
perturb the adiabatic population dynamics and prohibit the preparation of a
maximum coherence \cite{mph-stirap}.

In the present paper, we investigate the use the Stark Chirped Rapid
Adiabatic Passage (SCRAP) technique \cite{scrap,h-scrap,pop} to prepare
maximum coherence. In SCRAP, a pump laser couples a thermally populated
state (most likely the ground state) to an excited state and a second,
strong radiation pulse to induce a dynamic Stark shift. This Stark shift
serves to sweep the atomic transition frequency through resonance with the
pump laser frequency, mediating thereby an adiabatic passage of population
between two states. Provided the dynamic Stark shifts, induced by the second
laser are larger than the shifts, induced by the pump field, any
multi-photon transition may be used for the pump transition. For a
two-photon pump transition, it is therefore possible to create coherence
between a highly excited state and the ground state. If ultraviolet
radiation is used for the pump laser, coherence between states with energies
up to 10 eV may be efficiently created, permitting the generation of VUV
radiation well below 150 nm.

The potential of the nonlinear optics with maximum coherence has been
demonstrated for the regime of undepleted coherence and hence for an
undepleted pump field \cite{jain96}. We consider here a process of
difference-frequency mixing involving a two-photon transition $\left|
1\right\rangle -\left| 2\right\rangle $ resonantly excited by a strong pump
field with frequency $\omega _{1}$ and a dipole-allowed transition $\left|
2\right\rangle -\left| 3\right\rangle $ excited off-resonance by an
''idler'' field with frequency $\omega _{2}$ (Fig. 1). In this case, energy
for the generated field is taken only from the pump field, which at the same
time participates in the preparation of the coherence. Therefore, it will
unavoidably be depleted if considerable conversion efficiency is expected.
It is the aim of the present work to clarify, how the conversion proceeds
when the pump field is depleted, what fraction of the total energy of the
pump field may be transferred to the generated (and idler) field, and which
parameters are needed in the specific case of coherence preparation by
SCRAP. To this end, we solve the nonlinear propagation problem taking into
account the pump field depletion. The solution of such a nonlinear problem
is particularly challenging for pulses and is in general possible only
numerically. In order to obtain analytical solutions we apply the so-called
Hamiltonian approach \cite{mel79,kryz,kors02} which allows for a solution in
a wide range of physically relevant situations. An essence of this formalism
is to reduce a set of Maxwell propagation equation to canonical Hamilton
equations of classical mechanics, which admit several integrals of motion.
Additionally, this approach allows to analyze phase matching conditions
taking into account intensity-dependent (Kerr effect) contributions, which
are not present in the simple treatment of undepleted coherence. This
approach is especially useful under adiabatic conditions, i.e. when the
atoms are excited by the laser pulses in such a way that they remain in the
same instantaneous eigenstate of the interaction Hamiltonian during the
entire process. This is the case in the atomic superposition prepared by
SCRAP, as it is discussed here.

Obviously the large atomic coherence should be maintained for the duration
of the conversion process. Since the coherent superposition includes a
highly excited state, this requirement restricts the conversion to a regime
with laser pulses of duration shorter than the natural lifetimes in the
system. This is however not in contradiction with the adiabaticity of SCRAP.
The adiabatic approximation requires a slow rate of evolution as compared to
the frequency separation of the adiabatic eigenstates. This results usually
in a requirement for the product of the pulse duration and the Rabi
frequency of the radiation field to be much larger than unity. Thus, for
sufficiently intense fields, the process can be adiabatic even for short
pulses. In the present paper, we consider the frequency conversion of short
pulses and disregard spontaneous relaxation processes.

The paper is organized as follows. In Sec.II we discuss general features and
advantages of nonlinear frequency conversion processes with respect to
atomic coherence. The coherence is assumed to be undepleted. In Sec. III the
preparation of large atomic coherence by SCRAP is described. Sec. IV and the
Appendix outline the Hamiltonian approach in nonlinear optics. Making use of
this formalism we derive analytic solutions for frequency conversion
processes involving maximum coherence in Secs.V, VI. Section V is devoted to
solutions at small propagation distances in which the pump field is not
depleted. Conclusions of Sec. II are confirmed and specific phase matching
conditions for the SCRAP method are derived. In Sec. VI full analytical
solutions are obtained taking into account depletion of the driving field.
In Sec. VII we discuss the conditions for phase matching with respect to the
compensation of phase mismatch, induced by linear refraction and the Kerr
effect simultaneously. Finally, in Sec. VIII, we consider the
spatio-temporal dynamics of the generated radiation pulse as well as the
evolution of the total conversion efficiency. Conclusions are presented in
Sec. IX.


\section{Frequency conversion with undepleted atomic coherence}


\bigskip The (pulsed) e.m. field propagating in an ensemble of three-level
atoms (Fig. 1) is assumed to consist of three components with carrier
frequencies $\omega _{1},\omega _{2}$ and $\omega _{3}=2\omega _{1}-\omega
_{2}$:
\begin{equation}
E(z,t)=\sum_{j}\left( \mathcal{E}_{j}(z,t)\exp (-i(\omega _{j}t-\mathbf{k}%
_{j}\mathbf{r}))+c.c.\right) .  \label{E}
\end{equation}
Here $\left| \mathbf{k}_{j}\right| =n_{j}\omega _{j}/c$\ with the refractive
index $n_{j}$ at frequency $\omega _{j}$ describing refraction due to levels
outside the three-level system in Fig. 1. The radiation pulses are supposed
to be shorter than the relaxation times in the atomic system. The waves $%
\mathbf{k}_{2}$ and $\mathbf{k}_{3}$ propagate at small angles with respect
to the vector $\mathbf{k}_{1}$ (the $z$ axis).


\begin{figure}[th]
\centerline{\epsfig{file=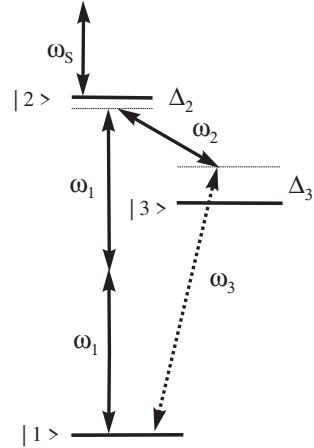,width=4.0 cm}} \vspace*{2ex}
\caption{Resonant four-wave mixing with maximum coherence between
$|1\rangle
$ and $|2\rangle $ adiabatically prepared by a strong drive $\protect\omega %
_{1}$ field and a ''Stark-shifting'' far-off-resonance $\protect\omega _{s}$
field.}
\label{system}
\end{figure}


In the approximation of slowly varying amplitudes and phases Maxwell's
propagation equations read in a moving frame
\begin{equation}
\frac{\partial \mathcal{E}_{j}}{\partial z}=i2\pi \frac{\omega _{j}}{c}%
\mathcal{P}_{j},  \label{Max1}
\end{equation}
where $\mathcal{E}_{j}$ and $\mathcal{P}_{j}$ ($j=1,2,3$) are functions of
the coordinate $z$ and the retarded time $\tau =t-z/c$.

$\mathcal{P}_{j}$ are the components of the medium polarization:
\begin{equation*}
P=\sum_{j}\Bigl(\mathcal{P}_{j}\exp (-i(\omega _{j}t-\mathbf{k}_{j}\mathbf{r}%
))+c.c.\Bigr).
\end{equation*}

The pulse at frequency $\omega _{s}$ (''ac Stark-shifting pulse'')
propagates along the $z$ axis and is far detuned from any atomic state.
Therefore, we assume that its intensity does not change along the
propagation path. In reality, the presence of the pulse at $\omega _{s}$
leads to generation of $2\omega _{1}\pm \omega _{s}$ frequency components.
However, an efficiency of their generation is much smaller than that for the
$\omega _{3}$ field due to the off-resonant character of the interaction of
the $\omega _{s}$ pulse with the medium. Moreover, the presence of these
(weak) components does not influence processes considered in the present
paper. Therefore, we disregard both the change of the $\omega _{s}$ pulse
intensity and the generation of $2\omega _{1}\pm \omega _{s}$ components.
The purpose of using the $\omega _{s}$ pulse is discussed later, in Sec.
III.C.

First we consider the situation of resonant nonlinear optics with maximum
coherence for constant amplitudes $c_{1}$ and $c_{2}$ of the states $%
|1\rangle $ and $|2\rangle $, prepared and maintained by a strong
pump field at $\omega _{1}$. The back-action of the atoms to this
field is disregarded and thus the corresponding coupling does not
need to be taken into account. In this regime the number of
photons in the preparatory field(s) must be much larger than the
number of atoms in the propagation volume. Additionally, the
prepared atomic coherence remains undepleted if the number of
generated $\omega _{3}$ photons is much smaller than the number of
the atoms in the relevant volume.

The components of the medium polarization can be expressed in terms of the
atomic probability amplitudes $c_{n}$ in levels $|1\rangle $, $|2\rangle $
and $|3\rangle $. After substitution into the Maxwell equation (\ref{Max1}),
one derives the field propagation equations \cite{jain96,har97o,kors02}:
\begin{eqnarray}
\frac{\partial \mathcal{E}_{2}^{\ast }}{\partial z} &=&i\frac{\pi N\omega
_{2}d_{2}^{2}}{\hbar c\Delta _{3}}\left| c_{2}\right| ^{2}\mathcal{E}%
_{2}^{\ast }+  \label{e21} \\
&&i\frac{\pi N\omega _{2}d_{2}d_{3}}{\hbar c\Delta _{3}}\rho _{12}e^{i\Delta
kz}\mathcal{E}_{3},  \notag \\
\frac{\partial \mathcal{E}_{3}}{\partial z} &=&i\frac{\pi N\omega
_{3}d_{3}^{2}}{\hbar c\Delta _{3}}\left| c_{1}\right| ^{2}\mathcal{E}_{3}+
\label{e31} \\
&&i\frac{\pi N\omega _{3}d_{2}d_{3}}{\hbar c\Delta _{3}}\rho
_{12}e^{-i\Delta kz}\mathcal{E}_{2}^{\ast },  \notag
\end{eqnarray}
where $N$ is the density of active atoms, $d_{2(3)}$ are the dipole moments
of transitions $|2\rangle \rightarrow |3\rangle $ ($|1\rangle \rightarrow
|3\rangle $), $\Delta _{3}$ is the frequency detuning indicated in Fig.~\ref
{system}:
\begin{equation}
\Delta _{3}=\omega _{3}-\omega _{31},  \label{d3}
\end{equation}
with $\omega _{nl}$ denoting the transition frequencies between the
corresponding levels. The atomic coherence between the states $|1\rangle $
and $|2\rangle $ is $\rho _{12}=\left| c_{1}c_{2}^{\ast }\right| $, and $%
\Delta k$ is the ''residual'' (background) phase mismatch:
\begin{equation}
\Delta k=k_{12}-k_{2}-k_{3},  \label{dk0}
\end{equation}
with $k_{j}$ $\left( j=2,3\right) $ being the projections of $\mathbf{k}_{j}$
on the $z$-axis. The wave vector $k_{12}$ of the atomic coherence $%
c_{1}^{\ast }c_{2}$ is related to the wave vector $k_{1}$ of the pump field
\cite{jain96,har97o}. In the case of two-photon excitation of the $|1\rangle
\rightarrow |2\rangle $ transition, considered in the present work, we have:
$k_{12}=2k_{1}$.

When deriving Eqs. (\ref{e21}) and (\ref{e31}), we disregarded a constant
phase of the atomic transition loop, and assumed large detuning $\left|
\Delta _{3}\right| \gg \Omega _{j}$, with $\Omega _{2}$ and $\Omega _{3}$
being the Rabi frequencies for transitions $\left| 2\right\rangle -\left|
3\right\rangle $ and $\left| 1\right\rangle -\left| 3\right\rangle $,
respectively:
\begin{equation}
\Omega _{j}=\frac{\left| d_{j}\mathcal{E}_{j}\right| }{2\hbar }.
\end{equation}

Equations (\ref{e21}) and (\ref{e31}) are linear differential equations,
which can easily be solved. We consider the case in which no $\mathcal{E}%
_{3} $ field is incident on the medium, $\mathcal{E}_{3}\left( z=0\right) =0$%
. Introducing the normalized intensity (photon flux)
\begin{equation}
\eta _{j}=\frac{{I_{j}}}{{\hbar \omega _{j}}}\equiv \frac{c\left| \mathcal{E}%
_{j}\right| ^{2}}{8\pi \hbar \omega _{j}}  \label{etas}
\end{equation}
and the coupling strength
\begin{equation}
\mu _{j}=\frac{2\pi \omega _{j}d_{j}^{2}}{\hbar c},  \label{mus}
\end{equation}
the solution of equations (\ref{e21}) and (\ref{e31}) reads:
\begin{align}
\eta _{2}(z)& =\eta _{20}\cosh ^{2}\left( \kappa z\sqrt{1-\left( \frac{%
\Delta k^{\prime }}{2\kappa }\right) ^{2}}\right)  \notag \\
& +\frac{\eta _{20}\left( \frac{\Delta k^{\prime }}{2\kappa }\right) ^{2}}{%
1-\left( \frac{\Delta k^{\prime }}{2\kappa }\right) ^{2}}\sinh ^{2}\left(
\kappa z\sqrt{1-\left( \frac{\Delta k^{\prime }}{2\kappa }\right) ^{2}}%
\right) ,  \label{sm2} \\
\eta _{3}(z)& =\frac{\eta _{20}}{1-\left( \frac{\Delta k^{\prime }}{2\kappa }%
\right) ^{2}}\,\,\sinh ^{2}\left( \kappa z\sqrt{1-\left( \frac{\Delta
k^{\prime }}{2\kappa }\right) ^{2}}\right) ,  \label{sm3}
\end{align}
where $\eta _{20}=\eta _{2}(z=0)$ is the photon flux at the entrance to the
medium. We have introduced the conversion coefficient $\kappa $:
\begin{equation}
\kappa =\frac{N}{2}\frac{\sqrt{\mu _{2}\mu _{3}}}{\Delta _{3}}\rho _{12},
\label{kap0}
\end{equation}
and
\begin{equation}
\Delta k^{\prime }=\Delta k+\frac{N}{2}\frac{\mu _{3}\left| c_{1}\right|
^{2}+\mu _{2}\left| c_{2}\right| ^{2}}{\Delta _{3}}  \label{dk-m}
\end{equation}
is the total phase mismatch, including the background value $\Delta k$ and
the contributions from resonant transitions $|1\rangle \rightarrow |3\rangle
$ and $|2\rangle \rightarrow |3\rangle $.

The solution Eqs. (\ref{sm2}) and (\ref{sm3}) shows that there is parametric
gain (exponential growth of intensity) for both $\omega _{2}$ and $\omega
_{3}$ waves with the rate $\kappa $ if the phase mismatch is compensated, $%
\Delta k^{\prime }\approx 0$. Since this rate is proportional to $\rho _{12}$%
, it is obviously advantageous to prepare atoms with large coherence on the $%
|1\rangle \rightarrow |2\rangle $ transition.

Phase match,
\begin{equation*}
\frac{2\Delta k}{N}\approx -\frac{\mu _{3}\left| c_{1}\right| ^{2}+\mu
_{2}\left| c_{2}\right| ^{2}}{\Delta _{3}},
\end{equation*}
can be achieved in several ways: (i) by tuning the wave vector $k_{21}$ of
the atomic coherence (e.g., via the detuning $\Delta _{2}$ of the pump
field, as in Ref. \cite{jain96,har97o}), (ii) by introducing a small angle
of the $\omega _{2}$ wave propagation direction from the $z$-axis, (iii) by
selecting the appropriate detuning $\Delta _{3}$, and/or (iv) by preparation
of atoms in a superposition with suitable amplitudes $c_{1},c_{2}$.

The resonant contributions to the phase mismatch, Eq. (\ref{dk-m}), are
usually the dominant ones over the residual value $\Delta k$. When phase
matching is not maintained, the quantity
\begin{equation*}
1-\left( \Delta k^{\prime }/2\kappa \right) ^{2}\approx -\frac{\left( \mu
_{3}\left| c_{1}\right| ^{2}-\mu _{2}\left| c_{2}\right| ^{2}\right) ^{2}}{%
4\mu _{2}\mu _{3}\rho _{12}^{2}}
\end{equation*}
is negative, i.e. no parametric gain but periodic change of intensity along
the propagation path takes place:
\begin{eqnarray}
\eta _{3}(z) &=&\eta _{20}\frac{4\mu _{2}\mu _{3}\rho _{12}^{2}}{\left( \mu
_{3}\left| c_{1}\right| ^{2}-\mu _{2}\left| c_{2}\right| ^{2}\right) ^{2}}%
\,\,  \notag \\
&&\times \sin ^{2}\left( \frac{N}{2}\frac{\left| \mu _{3}\left| c_{1}\right|
^{2}-\mu _{2}\left| c_{2}\right| ^{2}\right| }{2\Delta _{3}}z\right) .
\label{smb}
\end{eqnarray}
Moreover, in this regime, a substantial transfer of energy occurs for the
maximum coherence, $\rho _{12}\approx 1/2$, case, whereas it is small for
the regime of conventional nonlinear optics (weak excitation, $\left|
c_{2}\right| ^{2}\ll \left| c_{1}\right| ^{2}\approx 1$). The amount of
converted energy is larger for maximum coherence than for the conventional
nonlinear optics by a factor of the order of $\left| c_{2}\right| ^{-2}\gg 1$%
. We stress that the assumption of undepleted coherence assumes that the
number of generated $\omega _{3}$ photons is much smaller than the number of
photons in the preparatory field(s). Correspondingly, the \textit{total}
efficiency of energy conversion (from preparatory to generated fields) is
very small in this regime.

The regime of undepleted atomic coherence corresponds to the
classical picture of frequency mixing in which the atoms play the
role of a local oscillator (frequency $\omega _{21}$) with the
''probe'' $\omega _{2}$ field beating against it to produce the
difference (or sum-) frequency $\omega _{3}=\omega _{21}\pm \omega
_{2}$. Such a process is obviously more efficient for a strong
local oscillator, i.e., for large $\rho _{12}$. Thus, the
preparation of a large atomic coherence is favorable for frequency
conversion in atomic gases.

\section{Preparation of maximum coherence by SCRAP}

\subsection{Atomic parameters}

First we discuss the specific parameters to be considered in the coupling
scheme discussed here (Fig \ref{system}).

The Rabi frequencies of single-photon transitions $\Omega _{j}$ $\left(
j=2,3\right) $ are related to the photon flux $\eta _{j}$, Eq. (\ref{etas}),
via the coefficients $\mu _{j}$, Eq. (\ref{mus}), as
\begin{equation*}
\Omega _{j}=\sqrt{\mu _{j}\eta _{j}}.
\end{equation*}
The phase of the Rabi frequency $\Omega _{1}$ for a two-photon transition $%
|1\rangle \rightarrow |2\rangle $ is equal to $2\varphi _{1}$, and the
module is proportional to the intensity:
\begin{equation}
\Omega _{1}=\frac{1}{4\hbar }\alpha _{12}\left( \omega _{1}\right) \left|
\mathcal{E}_{1}\right| ^{2}\equiv \mu _{1}\eta _{1},  \label{Om1}
\end{equation}
where $\mu _{1}$ is the transition coupling constant, and $\alpha
_{nn^{\prime }}\left( \omega _{j}\right) $ is the matrix element of an
atomic polarizability tensor:
\begin{eqnarray}
\mu _{1} &=&\frac{2\pi \omega _{1}}{c}\alpha _{12}\left( \omega _{1}\right) ,
\label{mu1} \\
\hbar \alpha _{nn^{\prime }}\left( \omega _{j}\right) &=&\sum_{m}\left[
\frac{\left\langle n\right| d\left| m\right\rangle \left\langle m\right|
d\left| n^{\prime }\right\rangle }{\left( \omega _{m}-\omega _{n1}\right)
-\omega _{j}}\right.  \label{alphajj} \\
&&\left. +\frac{\left\langle n\right| d\left| m\right\rangle \left\langle
m\right| d\left| n^{\prime }\right\rangle }{\left( \omega _{m}-\omega
_{n^{\prime }1}\right) +\omega _{j}}\right] ,  \notag
\end{eqnarray}
with $\hbar \omega _{n1}$ being the energies of the resonant states $%
|n\rangle $ ($n=2,3$, with $\omega _{11}=0$), $\hbar \omega _{m}$ the
energies of the (virtual) states $|m\rangle $, and $\left\langle n\right|
d\left| m\right\rangle $ the dipole moment matrix elements for transitions $%
|n\rangle \rightarrow |m\rangle $ $\left( n=1,2,3\right) $.

The frequency detunings $\Delta _{n}$ $(n=2,3)$ include the ''static''
detuning $\delta _{n0}$, ac Stark shifts $\beta _{nj}\eta _{j}$ induced by
the $\omega _{j}$ $(j=1,2,3)$ fields, and the shifts $S_{n}=\beta _{ns}\eta
_{s}$ induced by an intense far-off-resonant ''SCRAP laser pulse'' with
frequency $\omega _{s}$ and photon flux $\eta _{s}$:
\begin{eqnarray}
\Delta _{n} &=&\delta _{n}+\sum_{j=1,2,3}\beta _{nj}\eta _{j},\qquad (n=2,3),
\label{dets} \\
\delta _{n} &=&\delta _{n0}+S_{n}, \\
\beta _{nj} &=&\frac{2\pi \omega _{j}}{c}\left( \alpha _{nn}\left( \omega
_{j}\right) -\alpha _{11}\left( \omega _{j}\right) \right) ,  \label{betas}
\\
\delta _{30} &=&\omega _{3}-\omega _{31},  \notag \\
\delta _{20} &=&2\omega _{1}-\omega _{21}.  \notag
\end{eqnarray}

It is important to note some essential relationships between the atomic
parameters used in the present work.

For atomic media the off-resonant (background) contributions to the
refractive index $n_{j}$ are expected to be of the order of \cite{boyd}:
\begin{equation}
n_{j}\approx 1+2\pi N\alpha _{11}\left( \omega _{j}\right) .  \label{n}
\end{equation}
The residual phase mismatch $\Delta k=2k_{1}-k_{2}-k_{3}$ is therefore of
the order of
\begin{equation}
\Delta k\approx \frac{2\pi N}{c}\left( 2\omega _{1}\alpha _{11}\left( \omega
_{1}\right) -\omega _{2}\alpha _{11}\left( \omega _{2}\right) -\omega
_{3}\alpha _{11}\left( \omega _{3}\right) \right) .  \label{dk-back}
\end{equation}
It follows then from Eqs. (\ref{mu1}), (\ref{alphajj}), (\ref{betas}), (\ref
{dk-back}) that the quantities:
\begin{equation*}
\mu _{1}\sim \beta _{nj}\sim \Delta k/N
\end{equation*}
are all of the same order of the magnitude.

Further, we have from Eqs. (\ref{mus}), (\ref{mu1}), (\ref{alphajj}):

\begin{equation}
\frac{\mu _{2,3}}{\mu _{1}}\sim \left| \left( \omega _{m}-\omega
_{n1}\right) \pm \omega _{j}\right| \gg \delta _{30},\delta _{20},\left|
\Omega _{j}\right|  \label{m2/m3}
\end{equation}
The last inequality is implied by the resonant three-level model of the
atom. The validity of this inequality justifies the use of the rotating wave
approximation.

Finally, we present values of the above constants for a real atomic scheme,
which can be used to drive the generation of short wavelength radiation. We
consider a coupling scheme in Kr with the states:\ $\left| 1\right\rangle
=4p^{6}$ $^{1}S$ (ground state), $\left| 2\right\rangle =4p^{5}$ $5p\left[
0,1/2\right] $ (94 093.7 $cm^{-1}$) and $\left| 3\right\rangle =4p^{5}$ $5s%
\left[ 1,1/2\right] $ (80 917.6 $cm^{-1}$). The two-photon transition
between the ground and the excited state in Kr is known as an efficient
transition for conventional four-wave mixing schemes. The scheme discussed
here has e.g. been used in experiments on VUV generation assisted by
electromagnetically induced transparency \cite{maran,dor00}. The pump field
at 212.55 nm excites the two-photon transition $\left| 1\right\rangle
-\left| 2\right\rangle $, the idler field at 759 nm is tuned near the
single-photon resonance of the transition $\left| 2\right\rangle -\left|
3\right\rangle $, and the field generated on the $\left| 3\right\rangle
-\left| 1\right\rangle $ transition has a wavelength of 123.6 nm. The
coupling strength of the single-photon transitions used in this scheme is : $%
\mu _{2}=3.507\times 10^{-2}$ $cm^{2}\times s^{-1}$ and $\mu
_{3}=0.441\times 10^{-2}$ $cm^{2}\times s^{-1}$. The coupling strength $\mu
_{1}$ of the two-photon transition and the ac Stark coefficients $\beta
_{2j} $ can be estimated as $\mu _{1}\approx 1.8\times 10^{-16}$ $cm^{2}$, $%
\beta _{21}\approx 3.7\times 10^{-17}$ $cm^{2}$, $\beta _{22}\approx
2.2\times 10^{-17}$ $cm^{2}$, $\beta _{23}\approx 6.4\times 10^{-17}$ $%
cm^{2} $. The residual phase mismatch for this scheme have been measured
\cite{dor00}, the value is: $\Delta k/N=4.8\times 10^{-17}$ $cm^{2}$.

\subsection{Interaction Hamiltonian, eigenvalue equation}

In rotating-wave approximation, the light-atom interaction Hamiltonian is
given by:
\begin{eqnarray}
\hat{H} &=&-\hbar \left[ \Delta _{2}\left| 2\right\rangle \left\langle
2\right| +\Delta _{3}\left| 3\right\rangle \left\langle 3\right| \right]
\label{Ham1} \\
&&-\hbar \Omega _{1}\left| 1\right\rangle \left\langle 2\right| +\hbar
\Omega _{2}e^{i\varphi }\left| 2\right\rangle \left\langle 3\right| +\hbar
\Omega _{3}\left| 1\right\rangle \left\langle 3\right| +H.c.,  \notag
\end{eqnarray}
where the multiphoton resonance condition
\begin{equation}
\omega _{3}=2\omega _{1}-\omega _{2}  \label{mph}
\end{equation}
has been used.

In the present paper, we consider adiabatic light-atom interaction
processes, i.e. the atomic system can be assumed to follow the evolution of
the instantaneous eigenstates. If, for example, the atomic system is at some
initial time $t_{0}$ in the nondegenerate eigenstate $\left| \psi
_{0}(t_{0})\right\rangle $ of the interaction Hamiltonian, i.e.
\begin{equation}
\hat{H}\left| \psi _{0}\right\rangle =\hbar \lambda _{0}\left| \psi
_{0}\right\rangle ,  \label{Eig1}
\end{equation}
(which is usually the ground state of the atoms), it will remain in this
state $\left| \psi _{0}\right\rangle $ at all times. Eq. (\ref{Eig1}) yields
the characteristic equation for the eigenvalues:
\begin{eqnarray}
&&\lambda _{0}\left( \Delta _{2}+\lambda _{0}\right) \left( \Delta
_{3}+\lambda _{0}\right) -\left( \Omega _{1}^{2}+\Omega _{2}^{2}+\Omega
_{3}^{2}\right) \lambda _{0}  \notag \\
&&\quad -\Omega _{1}^{2}\Delta _{3}-\Omega _{3}^{2}\Delta _{2}=-2\Omega
_{1}\Omega _{2}\Omega _{3}\cos \varphi ,  \label{Eig2}
\end{eqnarray}
where the relative phase $\varphi $ of the elm. waves is:
\begin{equation}
\varphi =2\varphi _{1}-\varphi _{2}-\varphi _{3}-\Delta kz,  \label{phase1}
\end{equation}
which includes the residual phase mismatch $\Delta k$.

\subsection{Preparation of maximum coherence by SCRAP procedure}

In what follows, we will concentrate on a regime of nonlinear optics with
large coherence between the states $\left| 1\right\rangle $ and $\left|
2\right\rangle $. To prepare such a coherence efficiently, we suggest to use
the Stark Chirped Rapid Adiabatic Passage (SCRAP) method \cite{scrap,h-scrap}%
. In the particular example of three-level system in Fig. 1, this procedure
can be realized when the idler $\omega _{2}$ field is far detuned from the
resonance with transition $\left| 2\right\rangle -\left| 3\right\rangle $
(i.e., the static detuning $\delta _{30}$ is much larger than other
parameters including all the Rabi frequencies and detuning $\Delta _{2}$).
In this case, the adiabatic (dressed) state that asymptotically connects to $%
\left| 1\right\rangle $ for $t\rightarrow -\infty $ is given by:
\begin{equation}
\left| \psi _{0}\right\rangle \approx \frac{\Omega _{1}}{\sqrt{\lambda
_{0}^{2}+\Omega _{1}^{2}}}\left| 1\right\rangle -\frac{\lambda _{0}}{\sqrt{%
\lambda _{0}^{2}+\Omega _{1}^{2}}}\left| 2\right\rangle ,  \label{psi0}
\end{equation}
with corresponding energy eigenvalue
\begin{equation}
\lambda _{0}\approx -\frac{1}{2}\Delta _{2}+\frac{1}{2}\sqrt{\Delta
_{2}^{2}+4\Omega _{1}^{2}}.  \label{lam0}
\end{equation}

As it is obvious from these formulae, population can be prepared from the
\textit{bare} state $\left| 1\right\rangle $ in the dressed state $\left|
\psi _{0}\right\rangle $, if at the beginning of the interaction $\Delta
_{2}\rightarrow +\infty $. The state $\left| \psi _{0}\right\rangle $ will
project completely onto the target bare state $\left| 2\right\rangle $ at
the end of the interaction, if $\Delta _{2}\rightarrow -\infty $. Thus all
the population can be transferred from the ground to the excited state via
the dressed state. For pulses with duration in the nanosecond range, this
transfer process can be implemented experimentally by sweeping the atomic
transition frequency with an additional laser pulse of frequency $\omega
_{s} $ (see Fig. 1) inducing dynamic Stark shifts (SCRAP). The pump and
Stark shifting laser pulses have to be delayed. Otherwise the effect of
rapid adiabatic passage will occur twice, once in the rising, second in the
falling wing of the Stark shifting laser.


\begin{figure}[th]
\centerline{\epsfig{file=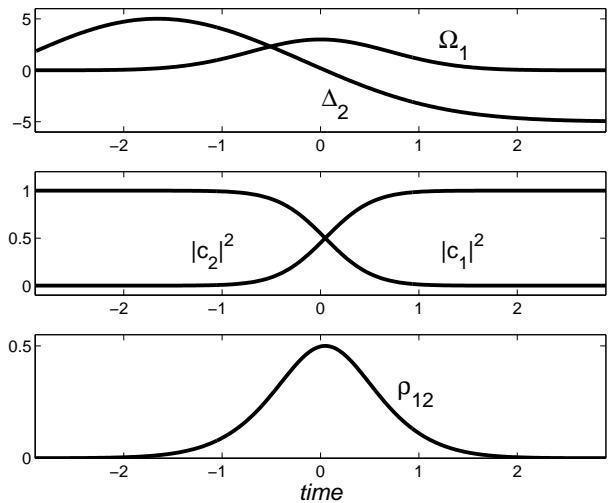,width=8.0 cm}} \vspace*{2ex}
\caption{SCRAP procedure producing complete population transfer
and a pulse of large atomic coherence. Time evolution of the
detuning $\Delta _{2}$ and the two-photon Rabi frequency $\Omega
_{1}$\ (top frame), the populations of the $\left| 1\right\rangle
$ and $\left| 2\right\rangle $ states (middle frame), and the
coherence $\protect\rho _{12}$ (bottom frame).} \label{SCRA1}
\end{figure}


Fig. \ref{SCRA1} shows the population and coherence dynamics induced in an
atomic system driven by laser pulses in SCRAP configuration. As described
above, population is transferred completely from the ground to the excited
state, as the atomic transition frequency is swept in the falling edge of
the Stark shifting laser pulse through resonance with the pump laser (middle
frame). The coherence $\rho _{12}$ induced in the system during the
interaction reaches a maximum of $1/2$ when the population is distributed
equally between the bare states $\left| 1\right\rangle $ and $\left|
2\right\rangle $. The system is prepared in maximum coherence. This happens,
however, only at one instant of time, which results in a pulse of large
atomic coherence $\rho _{12}\left( t\right) $. We note that the transient
large coherence $\rho _{12}$ occurs also when the pump and Stark shifting
laser pulses coincide. As we show later, however, such regime leads to quite
low overal conversion efficiency due to phase matching reasons.


\begin{figure}[th]
\centerline{\epsfig{file=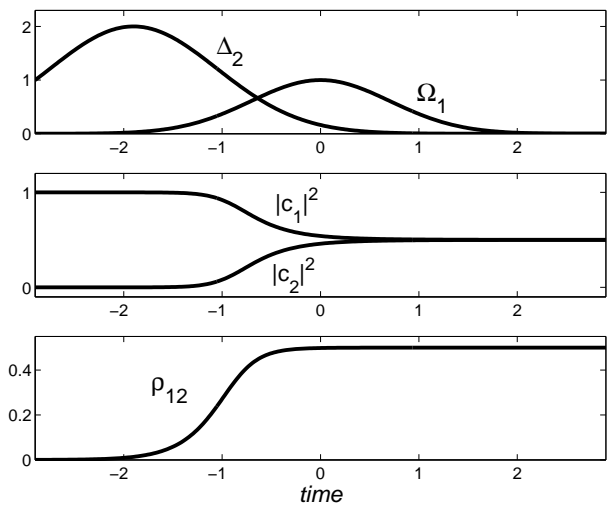,width=8.0 cm}} \vspace*{2ex}
\caption{''Half-SCRAP'' procedure leading to a permanent coherence
after the interaction. Time evolution of the detuning $\Delta
_{2}$ and the two-photon Rabi frequency $\Omega _{1}$\ (top
frame), the populations of the $\left| 1\right\rangle $ and
$\left| 2\right\rangle $ states (middle frame), and the coherence
$\protect\rho _{12}$ (bottom frame).} \label{SCRA2}
\end{figure}


While the coherence, induced by SCRAP, is not permanent, a slight
modification of the process permits the preparation of a long lasting
maximum coherence (so called half-SCRAP) \cite{h-scrap}. In this
configuration the pump laser is tuned to resonance, i.e. the static detuning
$\delta _{20}=0$. Fig. \ref{SCRA2} shows the population dynamics and the
coherence induced in this case. At earlier times we have $\Delta _{2}\gg
\Omega _{1}$ so that the state $\left| \psi _{0}\right\rangle $ coincides
with the ground state $\left| 1\right\rangle $. When $\Omega _{1}$
increases, the adiabatic state $\left| \psi _{0}\right\rangle $ evolves in a
superposition $\left| 1\right\rangle $ and $\left| 2\right\rangle $. At the
end of the interaction the situation $\Omega _{1}\gg \left| \Delta
_{2}\right| $ is reached, and the adiabatic state corresponds to ''maximum
coherence'': $\left| \psi _{0}\right\rangle =\left( \left| 1\right\rangle
-\left| 2\right\rangle \right) /\sqrt{2}$. This regime requires to fix the
static detuning $\delta _{20}$ to zero with sufficient accuracy, but it
establishes the enduring coherence needed for phase matching.

\section{Hamiltonian approach formalism}


In this work, we use an approach which does not require explicit expressions
for the atomic amplitudes \cite{mel79,kryz,kors02}.The main advantage of
this approach is the reduction of Maxwell propagation equations (\ref{Max1})
to the form of canonical Hamilton equations of classical mechanics involving
action and angle variables $J$ and $\varphi $ (see Appendix). Here $\varphi $
is the relative phase of e.m. waves, Eq. (\ref{phase1}), and the variable $%
J(z)$ characterizes the amount of energy exchange between the waves and has
the initial condition $J(z=0)=0$:
\begin{eqnarray}
\eta _{1}(z) &=&\eta _{10}-2J(z),  \notag \\
\eta _{2}(z) &=&\eta _{20}+J(z),  \label{etas2} \\
\eta _{3}(z) &=&\eta _{30}+J(z).  \notag
\end{eqnarray}

After some algebra (see Appendix), the Hamilton equations can be further
reduced to yield an implicit solution for $J(z)$:
\begin{equation}
\pm \frac{N}{2}z=\int\limits_{0}^{J}S(J^{\prime })\frac{dJ^{\prime }}{\sqrt{%
R\left( J^{\prime }\right) }},  \label{int}
\end{equation}
where both functions $R\left( J\right) $ and $S\left( J\right) $ are
polynomials in $J$.

For the generation of the $\omega _{3}$\ mode from vacuum: $\eta _{30}=0$,
the functions $R\left( J\right) $ and $S\left( J\right) $ take a form:
\begin{eqnarray}
R &=&4\mu _{1}^{2}\mu _{2}\mu _{3}J\left( \eta _{10}-J\right) ^{2}\left(
\eta _{20}+J\right)  \label{R} \\
&&-\left( A_{1}+A_{2}J+A_{3}J^{2}\right) ^{2}J^{2},  \notag \\
S &=&a_{0}+a_{1}J+a_{2}J^{2}.  \label{S}
\end{eqnarray}
As shown in the Appendix, cf. Eq. (\ref{refr}), the coefficients $A_{m}$ and
$a_{m}$ describe the linear and nonlinear refraction coefficients of the
medium.

Equation (\ref{int}) matches a one-dimensional finite motion of a pendulum
in an external potential. The allowed range of $J$, corresponding to the
region of classically allowed motion of the pendulum, lies between $zero$
and the \textit{smallest positive} root $J_{1}$ of the polynomial equation:
\begin{equation}
R\left( J\right) =0.  \label{root1}
\end{equation}
The second term in expression (\ref{R}) for $R\left( J\right) $ is never
positive, so the smallest positive root of the polynomial is bounded by $%
\eta _{10}$. This reflects the fact that the conversion process stops when
the energy of the pump field is entirely depleted. In order to reach this
limit and thus to attain maximum conversion efficiency, the second term in (%
\ref{R}) should be small, which corresponds to negligible phase mismatch :
\begin{equation*}
A_{1}+A_{2}J+A_{3}J^{2}\approx 0.
\end{equation*}
In order to see which values are required to approximately satisfy this
condition, we have to analyze the coefficients $A_{m}$.

The coherence preparation process requires large static detuning $\delta
_{30}$ and small ac Stark shift induced by the idler $\omega _{2}$ wave: $%
\mu _{2}\eta _{20}/\delta _{30}\ll \mu _{1}\eta _{10}$ (see discussion in
Sec. III.C). Taking into account the relation $\mu _{2}/\mu _{1}\gg \delta
_{30}$, Eq. (\ref{m2/m3}), the latter requirement restricts the intensity of
the $\omega _{2}$ wave: $\eta _{20}\ll \eta _{10}$. Since in the
down-conversion process considered here the energy is taken only from the $%
\omega _{1}$ wave, this condition does not impose a real limitation.

Under these conditions, the non-vanishing coefficients $A_{m}$ and
$a_{m}$ are given by
\begin{eqnarray}
A_{1} &\simeq &-q\delta _{30}\left( 2\lambda +\delta _{2}+\beta _{21}\eta
_{10}\right)  \label{A1} \\
&&-\mu _{2}\lambda -\mu _{3}\left( \lambda +\delta _{2}+\beta _{21}\eta
_{10}\right) ,  \notag \\
A_{2} &\simeq &q^{2}\delta _{30}+\mu _{2}q+\mu _{3}\left( q-\beta
_{22}-\beta _{23}+2\beta _{21}\right) ,  \label{A2} \\
a_{0} &\simeq &\delta _{30}\left( 2\lambda +\delta _{2}+\beta _{21}\eta
_{10}\right) ,  \label{a0} \\
a_{1} &\simeq &-\left( \mu _{2}+\mu _{3}\right) ,  \label{a2}
\end{eqnarray}
where
\begin{equation}
q=2\Delta k/N  \label{q}
\end{equation}
The eigenvalue $\lambda $ is a constant of motion which can thus be found
from Eq. (\ref{lamb}):
\begin{eqnarray}
\lambda &=&\lambda _{0}=-\frac{1}{2}\left( \delta _{2}+\beta _{21}\eta
_{10}\right)  \notag \\
&&+\frac{1}{2}\sqrt{\left( \delta _{2}+\beta _{21}\eta _{10}\right)
^{2}+4\mu _{1}^{2}\eta _{10}^{2}}.  \label{lambda}
\end{eqnarray}


\section{Solutions for undepleted pump field}

We first consider the solution of Eq. (\ref{int}) for small density-length
products $Nz$ in the case of low conversion efficiencies, i.e. the intensity
of the generated field is assumed to be much smaller than the intensity of
the pump $\omega _{1}$ wave:
\begin{equation*}
\eta _{3}=J\ll \eta _{10}.
\end{equation*}
We can then neglect the term $A_{2}J$ in expression (\ref{R}), and the term $%
a_{1}J$ in expression (\ref{S}). In this case the solution of Eq. (\ref{int}%
) has a simple form:
\begin{eqnarray}
\eta _{3}(z) &=&\frac{\eta _{20}}{1-\left( \frac{\Delta k^{\prime }}{2\kappa
}\right) ^{2}}\,\,\sinh ^{2}\left( \kappa z\sqrt{1-\left( \frac{\Delta
k^{\prime }}{2\kappa }\right) ^{2}}\right) ,  \label{solun} \\
&&  \notag \\
\kappa &=&\frac{N}{2}\frac{\sqrt{\mu _{2}\mu _{3}}}{\delta _{30}}\frac{\mu
_{1}\eta _{10}}{2\lambda +\delta _{2}+\beta _{21}\eta _{10}}, \\
&&  \notag \\
\Delta k^{\prime } &=&-\frac{N}{2}\frac{A_{1}}{a_{0}}  \notag \\
&=&\Delta k+\frac{N}{2}\frac{\mu _{3}\left( \lambda +\delta _{2}+\beta
_{21}\eta _{10}\right) +\mu _{2}\lambda }{\delta _{30}\left( 2\lambda
+\delta _{2}+\beta _{21}\eta _{10}\right) },
\end{eqnarray}
which is similar to Eq. (\ref{sm3}) obtained under the assumption of
constant probability amplitudes of the states $|1\rangle $ and $|2\rangle $
(cf. Sect. II).

The optimum conversion (parametric gain with large rate $\kappa $) occurs
when the phase mismatch $\Delta k^{\prime }$ is compensated. Taking into
account Eq. (\ref{lambda}) for $\lambda $, the condition for phase matching
reads:
\begin{eqnarray}
\frac{2}{N}\Delta k &=&-\frac{\mu _{3}+\mu _{2}}{2\delta _{30}}  \notag \\
&&-\frac{\mu _{3}-\mu _{2}}{2\delta _{30}}\frac{\left( \delta _{2}+\beta
_{21}\eta _{10}\right) }{\sqrt{\left( \delta _{2}+\beta _{21}\eta
_{10}\right) ^{2}+4\mu _{1}^{2}\eta _{10}^{2}}}.  \label{phm}
\end{eqnarray}
It is very important to recognize that the parameters $\delta _{2}$ and $%
\eta _{10}$ are time-dependent (pulsed) in the SCRAP process. Therefore, the
r.h.s. of Eq. (\ref{phm}) is time-dependent. Hence, it is impossible to
phase-match the generated and the pump waves for the duration of all stages
of the light-atom interaction process. This fact has a detrimental effect
for the full SCRAP procedure where the pulse of large coherence is produced
(see discussion in Sec. III.C and Fig. \ref{SCRA1}). However, in the
half-SCRAP case with the permanent large coherence (Fig. \ref{SCRA2}), there
are two relatively long time intervals, in which the phase matching
condition does not depend on time.

At the early stage of the SCRAP process, when $\delta _{2}\gg \mu _{1}\eta
_{10}$, the condition is:
\begin{equation}
\frac{2}{N}\Delta k\approx -\frac{\mu _{3}}{\delta _{30}},  \label{phm1}
\end{equation}
and the nonlinear conversion coefficient takes the form:
\begin{equation}
\kappa _{1}\approx \kappa _{0}\frac{\mu _{1}\eta _{10}}{\delta _{2}},
\label{kap1}
\end{equation}
with the ''maximum coherence'' conversion coefficient $\kappa _{0}$:
\begin{equation}
\kappa _{0}=\frac{N}{2}\frac{\sqrt{\mu _{2}\mu _{3}}}{\delta _{30}}.
\label{kap00}
\end{equation}
For later times, when the Rabi frequency of the pump field exceeds the
detuning: $\mu _{1}\eta _{10}\gg \delta _{2}$ and the adiabatic state
corresponds to the maximum coherence superposition, the phase matching
condition becomes:
\begin{equation}
\frac{2}{N}\Delta k=-\frac{\mu _{3}+\mu _{2}}{2\delta _{30}}-\frac{\mu
_{3}-\mu _{2}}{2\delta _{30}}\frac{\beta _{21}}{\sqrt{\beta _{21}^{2}+4\mu
_{1}^{2}}},  \label{phm2}
\end{equation}
and the conversion coefficient is
\begin{equation}
\kappa _{2}\approx \kappa _{0}\frac{\mu _{1}}{\sqrt{\beta _{21}^{2}+4\mu
_{1}^{2}}}.  \label{kap2}
\end{equation}

Phase matching according to Eqs. (\ref{phm1}), (\ref{phm2}) can be performed
by controlling the background mismatch $\Delta k$ through a suitable choice
of the small angle of the $\omega _{2}$ wave propagation direction from the $%
z$-axis. For radiation in the visible spectral range, detuning $\delta
_{30}\sim $100 GHz and atom densities $N\sim 10^{13}\div 10^{14}$ $cm^{-3}$,
the value of $\Delta k$ necessary to compensate the resonance refraction
contributions, Eqs. (\ref{phm1}), (\ref{phm2}), corresponds to an angle of $%
0.1-1$ $mrad$.

It is obvious from Eqs. (\ref{kap1}), (\ref{kap2}) that $\kappa _{2}\gg
\kappa _{1}$. Therefore, it is advantageous to drive the frequency
conversion process when a large coherence is established, thus to apply the
pulse at $\omega _{2}$ when $\mu _{1}\eta _{10}\gg \delta _{2}$ and to
choose parameters satisfying the condition Eq. (\ref{phm2}).

When phase matching is not maintained, the quantity $1-\left( \Delta
k^{\prime }/2\kappa \right) ^{2}$ is always negative ($\Delta k$ is much
smaller than the resonant contributions to $\Delta k^{\prime }$), and there
is no exponential growth but sinusoidal oscillations of the generated
intensity with respect to $Nz$. However, for $\mu _{1}\eta _{10}\gg \delta
_{2}$ (maximum coherence) the quantity $\left| 1-\left( \Delta k^{\prime
}/2\kappa \right) ^{2}\right| $ in the denominator of Eq. (\ref{solun}) is
of the order of unity, whereas for $\delta _{2}\gg \mu _{1}\eta _{10}$
(atoms are in the ground state) we have $\left| 1-\left( \Delta k^{\prime
}/2\kappa \right) ^{2}\right| \sim \left( \Delta k^{\prime }/2\kappa \right)
^{2}\gg 1$, and correspondingly, conversion is tiny.

These considerations, similar to those of Sect. II treating the nonlinear
conversion with fixed probability amplitudes, demonstrate once again that
large atomic coherence is preferable whatever the method of the preparation
might be.

Thus, in the case of good phase matching, we observe exponential growth of
the generated intensity. The question arises up to which values the
generated intensity will grow and what the limiting factors are? Since the
pump wave will be depleted, one may also expect that the preparation of
large atomic coherence will not be efficient anymore. By then it is not
clear how this will influence the frequency conversion process. In order to
answer these questions we need to solve the complete propagation problem
taking into account the depletion of the pump field.


\section{General solutions for resonant four-wave mixing}


The solution of the propagation equation (\ref{int}) is determined by the
roots of cubic equation (\ref{root1}), in particular, by their signs and
relation between their modules. Under the condition $\eta _{20}\ll \eta
_{10} $, the roots of (\ref{root1}) can be well approximated by:
\begin{eqnarray}
x_{1} &=&\frac{1-b_{1}}{1+b_{2}},  \label{x1} \\
x_{2} &=&\frac{1+b_{1}}{1-b_{2}},  \label{x2} \\
x_{3} &=&-\frac{\eta _{20}}{\eta _{10}}\frac{1}{1-b_{1}^{2}},  \label{x3}
\end{eqnarray}
where
\begin{equation*}
x_{j}=\frac{J_{j}}{\eta _{10}}.
\end{equation*}
The quantities $b_{1},b_{2}$:
\begin{eqnarray*}
b_{1} &=&\frac{A_{1}}{2\mu _{1}\eta _{10}\sqrt{\mu _{2}\mu _{3}}}, \\
b_{2} &=&\frac{A_{2}}{2\mu _{1}\sqrt{\mu _{2}\mu _{3}}}
\end{eqnarray*}
determine the phase mismatch induced by linear refraction and Kerr effect,
respectively.

Since $\eta _{20}/\eta _{10}\ll 1$ we have in most relevant cases: $\left|
x_{3}\right| \ll \left| x_{1}\right| ,\left| x_{2}\right| $.

Evaluation of the integral in Eq. (\ref{int}) gives the following general
dependence for $x(z)\equiv J\left( z\right) /\eta _{10}$ in implicit form:
\begin{eqnarray}
&&\pm \kappa ^{\prime }z+\chi _{0}=F\left[ \gamma \left( x\right) ,p\right]
\notag \\
&&\quad -\frac{a_{1}\eta _{10}}{a_{0}}r\left\{ F\left[ \gamma \left(
x\right) ,p\right] -d\Pi \left[ \gamma \left( x\right) ,n,p\right] \right\} ,
\label{sol00}
\end{eqnarray}
where $\chi _{0}$ is an integration constant, and $F\left( \gamma ,p\right) $
and $\Pi \left( \gamma ,n,p\right) $ are the elliptic integrals of the first
and third kind, respectively \cite{ell}. $\kappa ^{\prime }$ is the
nonlinear conversion coefficient defined as
\begin{equation}
\kappa ^{\prime }=\kappa _{0}\frac{\mu _{1}\eta _{10}\delta _{30}}{a_{0}}%
\sqrt{\left| 1-b_{1}^{2}\right| \left( 1+\frac{\left| x_{3}\right| }{\left|
x_{1}\right| }\right) }\frac{1}{s}.  \label{kappa}
\end{equation}
The parameters of the elliptic integrals $\gamma \left( x\right) ,n,p$ as
well as the factors $r,s,d$ depend on the signs of expressions $\left(
1-b_{1}^{2}\right) $ and $\left( 1-b_{2}^{2}\right) $.

Due to the condition $\eta _{20}/\eta _{10}\ll 1$, the expression (\ref
{sol00}) can be inverted to give the explicit solutions presented below.

\subsection{Compensation of both linear refraction and Kerr effect : $%
b_{1}^{2}<1$ and $b_{2}^{2}<1$}

For $b_{1}^{2}<1$ and $b_{2}^{2}<1$ the relevant parameters are:
\begin{eqnarray}
\gamma \left( x\right) &=&\arcsin \sqrt{\frac{x\left( x_{1}+\left|
x_{3}\right| \right) }{x_{1}\left( x+\left| x_{3}\right| \right) }},  \notag
\\
p &=&\sqrt{\frac{x_{1}\left( x_{2}+\left| x_{3}\right| \right) }{x_{2}\left(
x_{1}+\left| x_{3}\right| \right) }},\;n=\frac{x_{1}}{x_{1}+\left|
x_{3}\right| },  \label{par1} \\
d &=&1,\quad s=1,\quad r=\left| x_{3}\right| .  \notag
\end{eqnarray}

In this case, the solution is as follows:
\begin{equation}
x\left( z\right) =\frac{x_{1}\left| x_{3}\right| \text{sn}^{2}\left[ \kappa
^{\prime }z;p\right] }{\left| x_{3}\right| +x_{1}\text{cn}^{2}\left[ \kappa
^{\prime }z;p\right] },  \label{sol01}
\end{equation}
where sn$\left[ \kappa z;p\right] $ and cn$\left[ \kappa z;p\right] $ are
the Jacobi elliptic sine and cosine functions, respectively \cite{ell}.

In this case, the parameter $p$ is close to unity, so that sn$\left[ \kappa
z;p\right] \rightarrow \tanh \left( \kappa z\right) $ and cn$\left[ \kappa
z;p\right] \rightarrow $sech$\left( \kappa z\right) $. Thus, for small
density-length products $Nz$ such that $\kappa ^{\prime }z\ll \ln \left(
x_{1}/\left| x_{3}\right| \right) $ the solution (\ref{sol01}) is reduced to
\begin{equation*}
x(z)=x_{3}\,\,\sinh ^{2}\left( \kappa ^{\prime }z\right) ,
\end{equation*}
which coincides exactly with the solution obtained under the condition of
undepleted pump field, see Eq. (\ref{solun}).

\begin{figure}[th]
\centerline{\epsfig{file=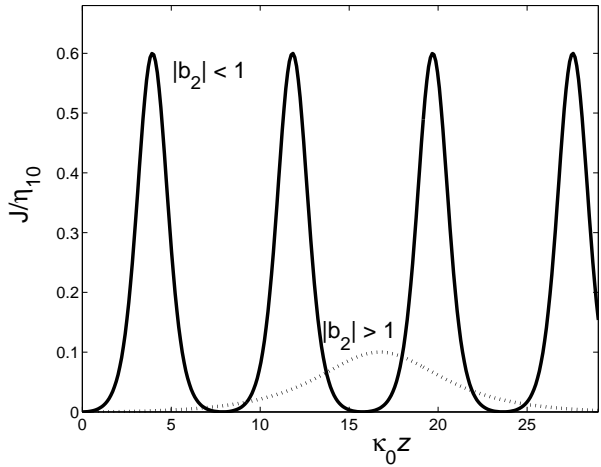,width=8 cm}} \vspace*{2ex}
\caption{Spatial evolution of $J\left( z\right) $ at given retarded time $%
\protect\tau $ in the case of compensated linear refraction. Parameters: $%
\protect\eta _{20}/\protect\eta _{10}=0.01$, $b_{1}=0.1$. The solid line
corresponds to Eq. (\ref{sol01}): $b_{2}=0.5$. The dotted line corresponds
to Eq. (\ref{sol02}): $b_{2}=8$, $s=5$.}
\label{conv1}
\end{figure}

For larger $Nz$, Eq. (\ref{sol01}) has to be used. The form of this solution
is shown by the solid line in Fig. \ref{conv1}. The maximum value of $x(z)$
attainable in this regime is given by $x_{1}$ corresponding to the intensity
of the generated $\omega _{3}$ wave given by :
\begin{equation*}
\left( \eta _{3}\right) _{max}=J_{max}=\eta _{10}\frac{1-b_{1}}{1+b_{2}},
\end{equation*}
which is of the order of $\eta _{10}$. Thus, almost complete conversion can
be achieved in this regime. This maximum value is reached at the distance $%
z=\left( \kappa ^{\prime }\right) ^{-1}K\left( p\right) $ with $K\left(
p\right) $ being a complete elliptic integral of the first kind, which can
be approximated for $p\approx 1-\left| x_{3}\right| \left(
x_{2}-x_{1}\right) /\left( x_{1}x_{2}\right) \rightarrow 1$ as
\begin{equation*}
K\left( p\right) \approx \left( 1/2\right) \ln \left( \frac{16x_{1}x_{2}}{%
\left( x_{2}-x_{1}\right) \left| x_{3}\right| }\right) .
\end{equation*}


\subsection{Compensation of linear refraction $b_{1}^{2}<1$, but large
Kerr-induced refraction $b_{2}^{2}>1$}

In this case, the parameters are as follows:
\begin{eqnarray}
\gamma \left( x\right) &=&\arcsin \sqrt{\frac{\left| x_{2}\right| \left(
x_{1}-x\right) }{x_{1}\left( x+\left| x_{2}\right| \right) }},  \notag \\
p &=&\sqrt{\frac{x_{1}\left( \left| x_{2}\right| -\left| x_{3}\right|
\right) }{\left| x_{2}\right| \left( x_{1}+\left| x_{3}\right| \right) }}%
,\;n=-\frac{x_{1}}{\left| x_{2}\right| },  \label{par2} \\
d &=&1-n,\quad s=1+\frac{a_{1}\eta _{10}}{a_{0}\sqrt{-n}},\quad r=\left|
x_{2}\right| .  \notag
\end{eqnarray}

The solution reads :
\begin{equation}
x\left( z\right) =\frac{x_{1}\left| x_{3}\right| \text{sn}^{2}\left[ \kappa
^{\prime }z;p\right] }{\left| x_{3}\right| +x_{1}\frac{2\left| x_{2}\right|
}{x_{1}+\left| x_{2}\right| }\text{cn}^{2}\left[ \kappa ^{\prime }z;p\right]
},  \label{sol02}
\end{equation}
The form of the solution is similar to the previous case, except for the
prefactor at the cn$^{2}\left[ \kappa ^{\prime }z;p\right] $ function in the
denominator. The spatial evolution of the generated intensity in this case
is plotted as a dotted line in Fig. \ref{conv1}.

We observe a parametric gain at the initial stage of the propagation, at $%
\kappa ^{\prime }z\ll \ln \left( \frac{x_{1}}{\left| x_{3}\right| }\frac{%
\left| x_{2}\right| }{x_{1}+\left| x_{2}\right| }\right) $:
\begin{equation}
x(z)=x_{3}\,\,\frac{x_{1}+\left| x_{2}\right| }{2\left| x_{2}\right| }\sinh
^{2}\left( \kappa ^{\prime }z\right) .  \label{ini02}
\end{equation}
The maximum of $x(z)$ from Eq. (\ref{sol02}) is again given by $x_{1}$.
However, for large ''Kerr coefficient'' $b_{2}\gg 1$, the value of $x_{1}$
corresponds to an intensity of the generated $\omega _{3}$ wave that is much
smaller than $\eta _{10}$:
\begin{equation}
\left( \eta _{3}\right) _{max}\sim \frac{\eta _{10}}{b_{2}}.
\label{max-eta2}
\end{equation}
It is also important to note that the conversion coefficient $\kappa
^{\prime }$ here is smaller than in the ''compensated case'' by a factor of $%
s\sim \left( \mu _{2,3}/\mu _{1}\delta _{30}\sqrt{-n}\right) \gg 1$.
Therefore, the conversion proceeds much slower, see Fig. \ref{conv1}.

\subsection{No compensation of linear refraction : $b_{1}^{2}>1$}

For $b_{1}^{2}>1$ and $b_{2}^{2}<1$ the elliptic integral parameters are:
\begin{eqnarray}
\gamma \left( x\right) &=&\arcsin \sqrt{\frac{x\left( x_{3}+\left|
x_{1}\right| \right) }{x_{3}\left( x+\left| x_{1}\right| \right) }},  \notag
\\
p &=&\sqrt{\frac{x_{3}\left( x_{2}+\left| x_{1}\right| \right) }{x_{2}\left(
x_{3}+\left| x_{1}\right| \right) }},\;n=\frac{x_{3}}{x_{3}+\left|
x_{1}\right| },  \label{par3} \\
d &=&1,\quad s=1,\quad r=\left| x_{1}\right| .  \notag
\end{eqnarray}

For $b_{1}^{2}>1$ and $b_{2}^{2}<1$ :
\begin{eqnarray}
\gamma \left( x\right) &=&\arcsin \sqrt{\frac{\left| x_{2}\right| \left(
x_{3}-x\right) }{x_{3}\left( x+\left| x_{2}\right| \right) }},  \notag \\
p &=&\sqrt{\frac{x_{3}\left( \left| x_{2}\right| -\left| x_{1}\right|
\right) }{\left| x_{2}\right| \left( x_{3}+\left| x_{1}\right| \right) }}%
,\;n=-\frac{x_{3}}{\left| x_{2}\right| },  \label{par4} \\
d &=&1-n,\quad s=1,\quad r=\left| x_{2}\right| .  \notag
\end{eqnarray}

In both cases $p\ll 1$ and $n\ll 1$. This permits a reduction of the
solution to the form:
\begin{equation}
x\left( z\right) =\left| x_{3}\right| \sin ^{2}\left( \kappa ^{\prime
}z\right) .  \label{sol03}
\end{equation}
The maximum of $x\left( z\right) $ is $\left| x_{3}\right| $, i.e. $%
J_{max}=\eta _{20}/\left| 1-b_{1}^{2}\right| $. Therefore, this solution
demonstrates the crucial influence of the phase mismatch induced by linear
refraction. If this contribution to the phase mismatch is not compensated,
the maximum intensity of the generated $\omega _{3}$ wave is always limited
by the input intensity $\eta _{20}$ of the idler wave.


\section{Compensation of phase mismatch}

As we have shown in the previous section, it is crucial to compensate the
mismatch induced by linear refraction $b_{1}^{2}<1$ in order to get large
conversion. Only then exponential gain occurs at the initial stage of the
process and the maximum generated intensity will be much larger than the
input intensity $\eta _{20}$ of the idler $\omega _{2}$ wave. At the same
time, it is desirable to make the Kerr-induced mismatch as small as possible.

\subsection{Compensation of the phase mismatch induced by linear refraction}

In general, the condition $b_{1}^{2}<1$ yields :
\begin{equation}
y_{0}-\frac{1}{\sqrt{1+d_{2}^{2}}}<y<y_{0}+\frac{1}{\sqrt{1+d_{2}^{2}}},
\label{b12<1}
\end{equation}
where the ''phase-matching-tuning parameter'' $y$ is:
\begin{equation}
y\equiv \frac{q\delta _{30}}{\sqrt{\mu _{2}\mu _{3}}}.  \label{y}
\end{equation}
The value of $y_{0}$:
\begin{equation}
y_{0}=-\frac{1-m}{2\sqrt{m}}\frac{d_{2}}{\sqrt{1+d_{2}^{2}}}-\frac{1+m}{2%
\sqrt{m}}\quad \left( y_{0}<0\right) ,  \label{y0}
\end{equation}
with notations
\begin{eqnarray}
m &=&\frac{\mu _{2}}{\mu _{3}},  \label{m} \\
d_{2} &=&\frac{\beta _{21}}{2\mu _{1}}+\frac{\delta _{2}}{2\mu _{1}\eta _{10}%
},  \label{d2}
\end{eqnarray}
corresponds to the condition given by Eq. (\ref{phm}) where $b_{1}=0$
(complete compensation of the linear refraction).

At the limits
\begin{equation}
y_{1,2}=y_{0}\mp 1/\sqrt{1+d_{2}^{2}}  \label{y12}
\end{equation}
of the desirable range of $y$, we have $b_{1}\rightarrow \pm 1$. From Eqs. (%
\ref{x1})-(\ref{x3}) we see that the root $x_{1}$ determining the maximum
conversion efficiency becomes very small at these values of $y$. Therefore,
it is not favorable to set the working point close to $y_{1,2}$.

We stress an important consequence of the inequality (\ref{b12<1}): For any $%
\delta _{2}$ and $\mu _{1}\eta _{10}$, the quantity $y<0$ as well as its
absolute value is of the order of one: $\left| y\right| \sim 1$ in the range
given by Eq. (\ref{b12<1}). Therefore, $q\sim \mu _{2,3}/\delta _{30}$ and
\begin{equation*}
\left| b_{2}\right| \sim \frac{\mu _{2,3}}{\mu _{1}\delta _{30}}\gg 1.
\end{equation*}
The Kerr-induced phase mismatch is large in the range of parameters ($q$ and
$\delta _{30}$) while linear refraction is compensated in this regime. Thus
it seems impossible to simultaneously compensate both contributions of the
phase mismatch.

\subsection{Compensation of the phase mismatch induced by the Kerr effect}

Condition $b_{2}^{2}<1$ yields :
\begin{equation}
\left| y\left( y+\frac{1+m}{\sqrt{m}}\right) \right| <\frac{2\mu _{1}\delta
_{30}}{\sqrt{\mu _{2}\mu _{3}}}.  \label{b22<1}
\end{equation}
Due to $2\mu _{1}\delta _{30}/\sqrt{\mu _{2}\mu _{3}}\ll 1$, the above
condition can be fulfilled by
\begin{equation}
\left| y\right| <\frac{2\mu _{1}\delta _{30}}{\mu _{2}+\mu _{3}}\quad \left(
\ll 1\right) ,
\end{equation}
i.e., by $y\approx y_{3}=0$ (or equivalently, by $q\approx 0$), and by
\begin{equation}
\left| y+\frac{1+m}{\sqrt{m}}\right| <\frac{2\mu _{1}\delta _{30}}{\sqrt{\mu
_{2}\mu _{3}}},
\end{equation}
that is by $y\approx y_{4}=-\left( 1+m\right) /\sqrt{m}$ [or equivalently,
by $q\delta _{30}\approx -\left( \mu _{2}+\mu _{3}\right) $, see Eq. (\ref
{phm2})] in a very small range $\pm 2\mu _{1}\delta _{30}/\left( \mu
_{2}+\mu _{3}\right) $. The values for $y_{3,4}$ are fixed by atomic
parameters and cannot be tuned.

Further, it is easy to show that
\begin{equation}
y_{4}\leq y_{1}<y_{2}<0.  \label{ym-y1}
\end{equation}

We see that the only possibility to satisfy both $b_{1}^{2}<1$ and $%
b_{2}^{2}<1$ is to make the value $y_{1}$ close to $y_{m}$ and to tune $y$
to the vicinity of $y_{1}$ (and $y_{m}$). However, as we have discussed in
the previous subsection, it is not favorable to work with $y$ close to $%
y_{1} $ since the root $x_{1}$, which determines the maximum conversion,
becomes very small. Moreover, the condition $y_{1}=y_{m}$ reduces to
\begin{equation}
d_{2}=\frac{\left( 1-m\right) }{2\sqrt{m}},  \label{cond-d2}
\end{equation}
what can be realized only at one instant of time, because $d_{2}$ is a
time-dependent function, see Eq. (\ref{d2}). Then, only a very narrow part
of the generated pulse may, in principle, be phase-matched to the pump
waves, and any small deviation from the above condition given by Eq. (\ref
{cond-d2}) will destroy the phase matching.


\section{Spatio-temporal evolution. Total conversion efficiency}

\begin{figure}[th]
\centerline{\epsfig{file=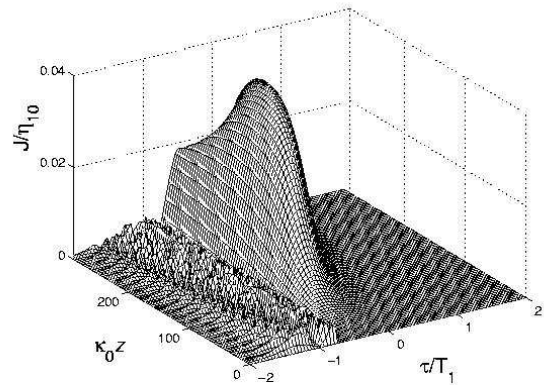,width=8 cm}} \vspace*{2ex}
\caption{Spatio-temporal evolution of generated intensity $J\left( z,\protect%
\tau \right) $ (normalized to the maximum of $\protect\eta _{10}$) for the
half-SCRAP preparation and linear refraction compensation at maximum
coherence according to Eq. (\ref{phm2}). The retarded time $\protect\tau $
is in units of the duration $T_{1}$ of the pump $\protect\omega _{1}$ pulse,
and\ the propagation distance $z$ is in units of conversion length $\protect%
\kappa _{0}^{-1}$ for the ideal maximum coherence case, Eq. (\ref{kap00}).
The parameters are $\protect\mu _{2}/\protect\mu _{3}=7.95$, $\protect\beta %
_{21}/2\protect\mu _{1}=0.1$ (Kr atoms), $\protect\mu _{2}/\left( 2\protect%
\mu _{1}\protect\delta _{30}\right) =20$. The temporal profile of the Stark,
pump and idler pulse is Gaussian with the following parameters: maximum of
detuning and pump Rabi frequency: $\protect\delta _{2}^{m}/\Omega
_{10}^{m}=2 $, static detuning $\protect\delta _{20}=0$, $\protect\eta %
_{20}^{m}/\protect\eta _{10}^{m}=0.005$, center of the Stark pulse $%
t_{s}/T_{1}=-1.5$, duration of the Stark pulse $T_{s}/T_{1}=1$, duration of
the $\protect\omega _{2}$ pulse $T_{2}/T_{1}=0.5$, delay of the $\protect%
\omega _{2}$ pulse $t_{2}/T_{1}=-1$.}
\label{fig3Da}
\end{figure}

\begin{figure}[th]
\centerline{\epsfig{file=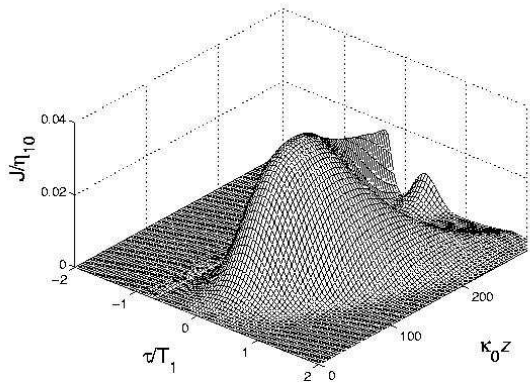,width=8 cm}} \vspace*{2ex}
\caption{Evolution of $J\left( z,\protect\tau \right) $ for the
half-SCRAP preparation and compensation of linear refraction at
maximum coherence by Eq. (\ref{phm2}). Parameters are the same as
in Fig. \ref{fig3Da} except the delay of the $\protect\omega _{2}$
pulse $t_{2}/T_{1}=0$. For better visibility, the time and length
axes have been reversed as compared to Fig. \ref{fig3Da}.}
\label{fig3Db}
\end{figure}

\begin{figure}[th]
\centerline{\epsfig{file=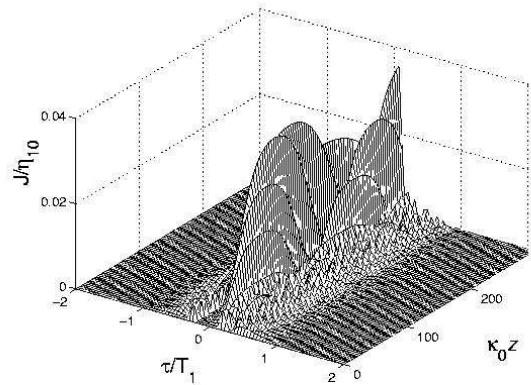,width=8 cm}} \vspace*{2ex}
\caption{Evolution of $J\left( z,\protect\tau \right) $ for the
temporally large coherence induced during the full SCRAP process.
Linear refraction is
compensated for maximum coherence by Eq. (\ref{phm2}). Parameters are: $%
\protect\delta _{2}^{m}/\Omega _{10}^{m}=10$, $\protect\delta _{20}/\Omega
_{10}^{m}=-5$, $t_{s}/T_{1}=-1.7$, $T_{s}/T_{1}=2$, $t_{2}/T_{1}=0$. Other
parameters are the same as in Fig. \ref{fig3Da}. For better visibility, the
time and length axes have been reversed as compared to Fig. \ref{fig3Da}.}
\label{fig3Dc}
\end{figure}

As we have seen before, there are different phase matching regimes at
different stages of the SCRAP process. If we choose to compensate the linear
refraction by satisfying the relationship (\ref{phm2}), then at the
beginning when $\delta _{2}\gg \mu _{1}\eta _{10}$ the linear refraction is
large: $\left| b_{1}\right| >1$. When the pump intensity reaches the values $%
\mu _{1}\eta _{10}$ $\gg \delta _{2}$ the mismatch is completely
compensated. Therefore, it is favorable to start the conversion process,
i.e., to apply $\omega _{2}$ pulse at time instants when $\mu _{1}\eta _{10}$
$\gg \delta _{2}$. However, the delay of $\omega _{2}$ pulse should not be
too large since the conversion process is also determined by the overlap
between the pump $\omega _{1}$ and the idler $\omega _{2}$ pulses. In order
to illustrate these processes, we show graphical representation of our
analytical results in Figs. \ref{fig3Da}, \ref{fig3Db}, \ref{fig3Dc}. Figs.
\ref{fig3Da} and \ref{fig3Db} demonstrate the evolution of the generated
intensity in the case of the permanent coherence preparation by the
half-SCRAP, and Fig. \ref{fig3Dc} - for the case of pulsed large atomic
coherence induced during the population transfer in the full SCRAP process.


In Fig. \ref{fig3Da}, with the idler $\omega _{2}$ pulse arriving before the
pump $\omega _{1}$ pulse, sinusoidal oscillations occure along the
propagation path\ for the early part of the generated pulse. This
corresponds to Eq. (\ref{sol03}) for the case of uncompensated phase
mismatch. As the pump intensity $\eta _{10}$ increases and the interaction
parameters get closer to the phase matching condition Eq. (\ref{phm2}) and $%
\left| b_{1}\right| $ becomes sufficiently small: $\left| b_{1}\right| <1$,
Eq. (\ref{sol02}) is applied. During this time interval intensities $J$ much
larger than $\eta _{20}$ do occur. We recall that $\eta _{20}$ is the
maximum intensity that can be obtained without elimination of linear
refraction. However, since the Kerr-induced mismatch is large, the maximum
generated intensity $\eta _{3}=J$ is still much smaller than $\eta _{10}$
(it is given by $\eta _{10}/b_{2},$ Eq. (\ref{max-eta2})), and the rate of
intensity growth is quite small.


Fig. \ref{fig3Db} shows the variation of the intensity $J$ with $\tau $ and $%
z$, when the phase matching takes place over the entire duration of the
generated pulse. We see from Figs. \ref{fig3Da}, \ref{fig3Db} that the
maximum of the generated pulse always coincides with the maximum of the pump
$\omega _{1}$ pulse and not with that of the idler $\omega _{2}$ pulse. This
fact directly follows from the physics of the down-conversion process in
which the $\omega _{2}$ field serves simply as a seed wave, while the energy
is taken only from the pump $\omega _{1}$ field. However, the temporal
overlap between the pump and the idler pulses also influences the conversion
process. Initially, there is exponential growth with $x(z)\sim x_{3}\sinh
^{2}\left( \kappa ^{\prime }z\right) \sim \eta _{20}/\eta _{10}$, Eq. (\ref
{ini02}). Thus the generated intensity is determined mainly by the idler
field: $\eta _{3}\sim \eta _{20}\sinh ^{2}\left( \kappa ^{\prime }z\right) $%
. Later, the maximum of the generated pulse is shifted towards the maximum
of the $\omega _{1}$ pulse. This dynamics leads, in general, to a temporal
modulation of the generated pulse. The best conditions are obtained when the
maxima of the pump $\omega _{1}$ and the idler $\omega _{2}$ pulses coincide
(Fig. \ref{fig3Db}). In this case, conversion proceeds more or less
homogeneously and the temporal shape of the generated pulse is smooth at all
propagation distances.


When the large atomic coherence persists only in transient during the full
SCRAP process, efficient generation occurs during a short time interval
around the maximum of the pump pulse (see Fig. \ref{fig3Dc}). This interval
is determined by the range where the phase matching, $\left| b_{1}\right| <1$
, occurs.

\begin{figure}[th]
\centerline{\epsfig{file=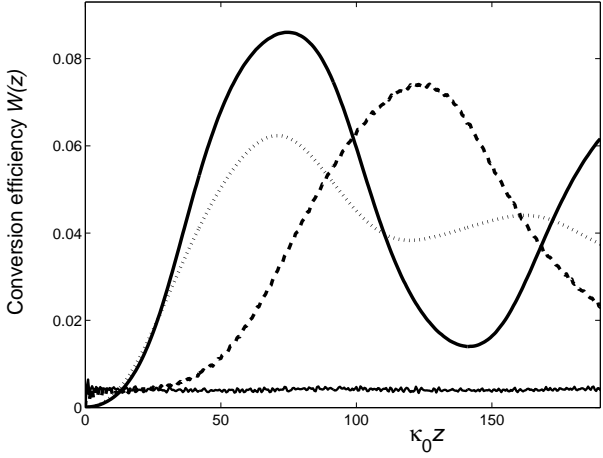,width=8 cm}} \vspace*{2ex}
\caption{Spatial dependence of the total conversion efficiency
$W$, Eq. (\ref {W}), for different delays of the $\protect\omega
_{2}$ pulse for the half-SCRAP coherence preparation: solid line -
$t_{2}/T_{1}=0$, dotted line - $t_{2}/T_{1}=+1$, dashed line -
$t_{2}/T_{1}=-1$. The values for all other parameters are the same
as in Fig. \ref{fig3Da}. The lower thin solid line corresponds to
the transient large coherence induced during the full SCRAP
process, Fig. \ref{fig3Dc}.} \label{eff1}
\end{figure}

As we see, the temporal shape of the generated pulse is in general quite
complicated. That is the output pulse is not transform-limited. An important
quantity to characterize the conversion process is the total energy
conversion efficiency, defined as
\begin{equation}
W(z)\equiv \frac{\displaystyle{\int \mathrm{d}t\,\omega _{3}\eta _{3}(z,t)}}{%
\displaystyle{\ \int \mathrm{d}t\,\omega _{1}\eta _{10}(z,t)}}.  \label{W}
\end{equation}
The evolution of $W(z)$ is shown in Fig. \ref{eff1} for three different
delays of the idler pulse in the case of permanent coherence. The largest
conversion efficiency is obtained for coinciding $\omega _{1}$ and $\omega
_{2}$ pulses. Moreover, the maximum of $W(z)$ occurs at propagation
distances smaller than that for the case of delayed pulses. However, the
total conversion efficiency is not substantially different for different
delays because the maximum of $W(z)$ is determined mainly by $\eta
_{10}^{m}/b_{2}$ , i.e., by the parameters of the pump pulse. The thin solid
line in Fig. \ref{eff1} displays the conversion efficiency in the case of
pulsed large coherence. As expected, the efficiency is much smaller than in
the ''permanent $\rho _{12}$'' case because of two reasons. First, the
nonlinear conversion coefficient $\kappa $, Eq. (\ref{kap0}), is large in
the transient regime during a short time slot. Second, more important, phase
matching, as discussed above, can be achieved only in an even shorter time
interval.


\section{Conclusions}

We have discussed the analytic solutions of a four-wave mixing
process involving preparation of an atomic system driven to
maximum coherence by the technique of Stark chirped rapid
adiabatic passage (SCRAP). The maximum coherence permits
conversion efficiencies, exceeding the case of conventional
nonlinear optics by a large factor. However, the conversion
efficiency does not reach unity, because of phase mismatch due to
the linear and the intensity-dependent index of refraction. It is
practically impossible to compensate both parts simultaneously.
Thus, the conversion efficiency gets maximum when the linear part
of the phase mismatch is reduced to zero. This can be done by the
controlling the residual phase mismatch $\Delta k$ through the
buffer gas or non-collinear pulse propagation. Unfortunately, the
Kerr-induced mismatch is large and limits the rate and the maximum
achievable efficiency of the conversion. Still the maximum atomic
coherence, prepared by SCRAP, permits efficient generation of
strong short-wavelength radiation. In principle, the generated
radiation is broadly tunable - the detuning $\delta _{30}$ may be
changed provided it is still larger than Rabi frequencies and ac
Stark shifts. However, the phase matching condition Eq.
(\ref{phm2}) has to be satisfied. Therefore, tuning of $\delta
_{30}$ requires modified compensation of the residual phase
mismatch $\Delta k$.

\section{Acknowledgments}

We acknowledge support from the European Union, through the RTN COCOMO
contract number HPRN-CT-1999-000129, the Deutsche Forschungsgemeinschaft
(DFG) as well as the German-Israeli Foundation (GIF), contract number
I-644-118.5/1999. The work of E.A.K. was supported by the Alexander von
Humboldt Foundation. We would like to thank M. Fleischhauer for many useful
discussions.

\begin{appendix}

\section{Hamiltonian approach}




Here we present an outline of the Hamiltonian approach in nonlinear
optics \cite{mel79,kryz,kors02}. This approach is based on the
representation of the medium polarization $P$ as a partial derivative of the
time-averaged free energy density of a dielectric with respect to the
electric field strength $E$ \cite{LL}:
\begin{equation}
P=-\left\langle N\frac{\partial \hat{H}}{\partial E}\right\rangle ,
\label{Pol2}
\end{equation}
where $\left\langle ...\right\rangle $ denotes quantum-mechanical averaging,
and $\hat{H}$ is the single-atom interaction Hamiltonian. With the field
given by Eq. (\ref{E}), we write:
\begin{equation}
P=-\left\langle N\sum_{j}\frac{\partial \hat{H}}{\partial \mathcal{E}%
_{j}^{\ast }}\exp (-i(\omega _{j}t-k_{j}z))+c.c.\right\rangle ,  \label{Pol3}
\end{equation}
thus the propagation equation (\ref{Max1}) becomes:
\begin{equation}
\frac{\partial \mathcal{E}_{j}}{\partial z}=-i2\pi \frac{\omega _{j}}{c}%
N\left\langle \frac{\partial \hat{H}}{\partial \mathcal{E}_{j}^{\ast }}%
\right\rangle .  \label{Max2}
\end{equation}
When the atomic system adiabatically follows the instantaneous
eigenstate $\left| \psi _{0}\right\rangle $, we find
\[
\left\langle \frac{\partial \hat{H}}{\partial \mathcal{E}_{j}^{\ast }}%
\right\rangle =\left\langle \psi _{0}\right| \frac{\partial \hat{H}}{%
\partial \mathcal{E}_{j}^{\ast }}\left| \psi _{0}\right\rangle =\hbar \frac{%
\partial \lambda _{0}}{\partial \mathcal{E}_{j}^{\ast }}.
\]
Hence the propagation equation can be written as:
\begin{equation}
\frac{\partial \mathcal{E}_{j}}{\partial z}=-i2\pi \frac{\hbar \omega _{j}}{c%
}N\frac{\partial \lambda _{0}}{\partial \mathcal{E}_{j}^{\ast }}.
\label{Max3}
\end{equation}
In the following discussion it is useful to express the field amplitude $\mathcal{E}%
_{j}$ in terms of photon flux $\eta _{j}$, Eq. (\ref{etas}), and phase $%
\varphi _{j}$. Separating the real and imaginary parts, we find from Eq. (%
\ref{Max3}):
\begin{eqnarray}
\frac{\partial \eta _{j}}{\partial z} &=&-\frac{\partial \mathcal{H}^{\prime
}}{\partial \varphi _{j}},  \label{Can1} \\
\frac{\partial \varphi _{j}}{\partial z} &=&\frac{\partial \mathcal{H}%
^{\prime }}{\partial \eta _{j}}.  \nonumber
\end{eqnarray}
These equations have the form of Hamilton equations of classical canonical
mechanics with action and angle variables $\eta _{j}$, and $\varphi _{j}$,
''time'' $z$, and the Hamiltonian function $\mathcal{H}^{\prime }=\frac{1}{2}%
N\lambda _{0}$.




One can see from the eigenvalue equation (\ref{Eig2}) that $\lambda _{0}$ and,
hence $\mathcal{H}^{\prime }$, depend on the field phases $\varphi _{j}$
only through the relative phase $\varphi $. Therefore, we have:
\begin{equation}
\frac{\partial \mathcal{H}^{\prime }}{\partial \varphi _{1}}=-2\frac{%
\partial \mathcal{H}^{\prime }}{\partial \varphi _{2}}=-2\frac{\partial
\mathcal{H}^{\prime }}{\partial \varphi _{3}}\left( =2\frac{\partial
\mathcal{H}^{\prime }}{\partial \varphi }\right) .  \label{phc}
\end{equation}
An immediate consequence of this symmetry of $\mathcal{H}^{\prime }$ is the
existence of constants of motion. Substituting the above equations (\ref{phc}%
) into the first line of Eqs. (\ref{Can1}) yields the well-known Manley-Rowe
relations \cite{boyd}:
\begin{equation}
\frac{\partial \eta _{1}}{\partial z}=-2\frac{\partial \eta _{2}}{\partial z}%
=-2\frac{\partial \eta _{3}}{\partial z},  \label{MR}
\end{equation}
which correspond to two independent constants of motion:
\begin{eqnarray}
\eta _{1}+2\eta _{3} &=&\eta _{10}+2\eta _{30},  \label{const1} \\
\eta _{1}+2\eta _{2} &=&\eta _{10}+2\eta _{20}.  \nonumber
\end{eqnarray}
Here $\eta _{j0}=\eta _{j}(z=0)$ are the photon flux values at the entrance
to the medium. Taking into account the multiphoton resonance condition (\ref
{mph}), one finds furthermore that the total intensity of the elm. fields is
conserved: $I_{1}+I_{2}+I_{3}=const(z)$.
The relations Eq. (\ref{const1}) enable us to re-write $\eta _{j}$ as:
\begin{eqnarray}
\eta _{1}(z) &=&\eta _{10}-2J(z),  \nonumber \\
\eta _{2}(z) &=&\eta _{20}+J(z),  \label{etas3} \\
\eta _{3}(z) &=&\eta _{30}+J(z).  \nonumber
\end{eqnarray}
The function $J(z)$ characterizes the amount of energy exchange between the
waves and has the initial condition $J(z=0)=0$.

Thus the original problem with six amplitude and phase variables can be
reduced to two variables $J$ and $\varphi $ by a canonical transformation.
This leads to
\begin{eqnarray}
\frac{\partial J}{\partial z} &=&-\frac{\partial \mathcal{H}}{\partial
\varphi },  \label{Can2J} \\
\frac{\partial \varphi }{\partial z} &=&\frac{\partial \mathcal{H}}{\partial
J},  \label{Can2phi}
\end{eqnarray}
with new Hamiltonian function
\begin{equation}
\mathcal{H}=\frac{1}{2}N\lambda _{0}+\Delta kJ\equiv \frac{1}{2}N\lambda .
\label{Ham2}
\end{equation}
As can be seen from Eqs. (\ref{Ham2}) and (\ref{Eig2}), $\mathcal{H}$ (or $%
\lambda $) does not depend on the coordinate $z$ explicitly. Therefore, $%
\mathcal{H}$ (or $\lambda $) is a fourth constant of motion expressing the
conservation of the energy density of the medium with respect to $z$.




To solve the remaining two equations of motion for $J(z)$ and $\varphi (z)$,
the Rabi-frequencies $\Omega _{j}$ are expressed in terms of $\eta _{j0}$
and $J$, and the characteristic equation (\ref{Eig2}) is written in the form
\begin{equation}
G(\lambda ,J)=g(J)\cos \varphi .  \label{Eig3}
\end{equation}
Differentiating both sides with respect to $\varphi $ yields
\[
\frac{\partial G}{\partial \varphi }=\frac{\partial G}{\partial \lambda }%
\frac{\partial \lambda }{\partial \varphi }=-g\sin \varphi =\pm \sqrt{%
g^{2}-G^{2}}.
\]
Substituting this relation into Eq. (\ref{Can2J}), we find:
\begin{equation}
\frac{\partial J}{\partial z}=\pm \frac{N}{2}\frac{\sqrt{g^{2}-G^{2}}}{%
\partial G/\partial \lambda }.  \label{fin}
\end{equation}
The choice of the sign in Eq. (\ref{fin}) depends on the sign of $\sin
\varphi $ at $z=0$. Integration of Eq. (\ref{fin}) gives an implicit
solution for $J(z)$:
\begin{equation}
\pm \frac{N}{2}z=\int\limits_{0}^{J}\frac{\partial G(J^{\prime })}{\partial
\lambda }\frac{dJ^{\prime }}{\sqrt{g^{2}\left( J^{\prime }\right)
-G^{2}\left( J^{\prime }\right) }}.  \label{int2}
\end{equation}
Both functions $g^{2}-G^{2}$ and $\partial G/\partial \lambda $ are
polynomials in $J$:
\begin{eqnarray}
g &=&-2\mu _{1}\sqrt{\mu _{2}\mu _{3}}  \label{f} \\
&&\times \sqrt{\left( \eta _{10}-2J\right) ^{2}\left( \eta _{20}+J\right)
\left( \eta _{30}+J\right) },  \nonumber \\
G &=&G_{0}+\sum_{m=1}^{3}A_{m}J^{m},  \label{F} \\
\frac{\partial G}{\partial \lambda } &=&\sum_{m=0}^{2}a_{m}J^{m}.  \label{dF}
\end{eqnarray}
Therefore, equation (\ref{fin}) describes a one-dimensional finite motion of
a pendulum in an external potential. The solution is in general given by
some combination of elliptic functions \cite{ell} with parameters determined
mainly by the roots $J_{n}$ of the polynomial equation:
\begin{equation}
g^{2}\left( J\right) -G^{2}\left( J\right) =0.  \label{root2}
\end{equation}
The eigenvalue $\lambda $ is a constant of motion (cf. Eq. (\ref{Ham2})),
and can thus be found from the characteristic equation (\ref{Eig3}) with
parameters taken at the medium entrance $z=0$:
\begin{equation}
G_{0}\left( \lambda \right) =g(z=0)\cos \varphi (z=0).  \label{lamb}
\end{equation}
Thus, we have reduced the propagation problem to solving two algebraic
equations: (\ref{lamb}) for $\lambda $ and (\ref{root2}) for the roots $%
J_{n} $. If this can be done explicitly, the Hamiltonian method
provides an analytical solution to the propagation problem. But
even if an explicit solution is not possible, it considerably
simplifies numerical calculations.
The physical meaning of the
coefficients $A_{m}$ and $a_{m}$ can be drawn by considering the
canonical equation (\ref{Can2phi}) for the relative phase:
\begin{equation}
\frac{\partial \varphi }{\partial z}=\frac{N}{2}\frac{\partial \lambda }{%
\partial J}=\frac{N}{2}\frac{\partial G/\partial J}{\partial G/\partial
\lambda }=\frac{N}{2}\frac{A_{1}+2A_{2}J+3A_{3}J^{2}}{a_{0}+a_{1}J+a_{2}J^{2}%
}.  \label{refr}
\end{equation}
One recognizes that the $A_{m}$ and $a_{m}$ describe the linear and
nonlinear refraction coefficients of the medium. E.g. if $J$ is sufficiently
small, the first term $NA_{1}/2a_{0}$ on the right-hand side of Eq. (\ref
{refr}) can be identified with the phase mismatch induced by the linear
refraction, including both contributions from the three-level interaction
and the residual mismatch $\Delta k$. The second term $\left(
N/2a_{0}\right) \left( 2A_{2}-a_{1}A_{1}/a_{0}\right) J$ in the expansion
over $J$ corresponds to the phase mismatch due to Kerr effect, and the next
terms are responsible for the higher-order contributions.

\end{appendix}


\end{document}